\documentclass[fleqn,10pt,a4paper]{article}
\usepackage{amsmath,amssymb,bm}
\usepackage{cite,epsfig,color}
\usepackage{fancyhdr}

\title{Geometrical description of non-linear electrostatic oscillations in
  relativistic thermal plasmas}
\author{DA Burton and A Noble\\Physics Department, Lancaster
  University, LA1 4YB, UK  \\ \& The Cockcroft Institute, UK}

\begin{document}
\maketitle
\begin{abstract}
We develop a method for investigating the relationship between the
shape of a 1-particle distribution and non-linear electrostatic oscillations in a
collisionless plasma, incorporating transverse thermal motion. A
general expression is found for the maximum sustainable electric
field, and is evaluated for a particular highly anisotropic distribution.
\end{abstract}
\section*{Introduction}
High-power lasers and plasmas may be
used to accelerate electrons by electric fields that are
orders of magnitude greater than those achievable using conventional
methods~\cite{tajima:1979}.
An intense laser pulse is used to drive a wave in an underdense
plasma and, for sufficiently large fields, non-linearities lead to
collapse of the wave structure (``wave-breaking'') due to
sufficiently large numbers of electrons becoming trapped in the
wave.

Hydrodynamic investigations of wave-breaking were first undertaken for cold
plasmas~\cite{akhiezer:1956, dawson:1959} and thermal effects were
later included in non-relativistic~\cite{coffey:1971} and relativistic
contexts~\cite{katsouleas:1988, rosenzweig:1988, schroeder:2005}
(see~\cite{trines:2006}
for a discussion of the numerous approaches). However, it is
clear that the value of the electric field at which the wave
breaks (the electric field's ``wave-breaking limit'')
is highly sensitive to the details of the hydrodynamic model.  

Plasmas dominated by collisions are described by a pressure tensor that does
not deviate far from isotropy, whereas an intense and ultrashort laser
pulse propagating through an underdense plasma will drive the plasma
anisotropically over typical acceleration timescales. Thus, it is
important to accommodate 3-dimensionality and allow for anisotropy
when investigating wave-breaking limits. The sensitivity of the
wave-breaking limit to the details of the plasma model suggests that 
it could depend on the anisotropy of the pressure tensor.
 
One method for investigating the wave-breaking limit of a
collisionless anisotropic
plasma is to employ the warm plasma closure of velocity moments of the
1-particle distribution $f$ satisfying the Vlasov-Maxwell
equations~\cite{schroeder:2005}. Successive order moments of the Vlasov
equation induce an infinite hierarchy of field
equations for the velocity moments of $f$ and at each finite order the
number of unknowns is greater than the number of field
equations. The warm plasma closure scheme sets the number of
unknowns equal to the number of field equations by assuming that
the terms containing the third order centred moment are negligible relative to
those including second, first and zeroth order centred moments.

Our aim is to uncover the relationship between wave-breaking and the
shape of $f$. In general, the detailed structure of $f$ cannot be reconstructed from a
few low-order moments so we adopt a different approach based on a
particular class of piecewise constant $1$-particle distributions.
Our choice of distribution, although somewhat artificial,
reduces the Vlasov equation to that of a boundary in the unit
hyperboloid bundle over spacetime. Combining the equation for the boundary with
the Maxwell equations yields an integral for the wave-breaking limit
in terms of the shape of the boundary.

Our approach may be considered
as a multi-dimensional generalization of the 
1-dimensional relativistic ``waterbag'' model employed in~\cite{katsouleas:1988}.
\section{Vlasov-Maxwell equations}
The brief summary of the Vlasov-Maxwell
equations given below establishes our conventions.
Further discussion of relativistic kinetic theory may
be found in, for example, ~\cite{ehlers:1971, degroot:1980}.
We employ the Einstein summation
convention throughout and units are used in which the speed of light
$c=1$ and the permittivity of the vacuum $\varepsilon_0=1$. Lowercase
Latin indices $a,b,c$ run over $0,1,2,3$.
\subsection*{{ Preliminary considerations}}
Let $(x^a)$ be an inertial coordinate system on Minkowski spacetime
$(\mathcal{M},g)$ where $x^0$ is the proper
time of observers at fixed Cartesian coordinates $(x^1,x^2,x^3)$ in
the laboratory. The metric tensor $g$ has the form
\begin{equation}
g = \eta_{ab}\, dx^a \otimes dx^b
\end{equation}
where 
\begin{equation}
\eta_{ab} =
\begin{cases}
&-1\text{   if $a=b=0$}\\
&1\text{    if $a=b\neq 0$}\\
&0\text{    if $a \neq b$}
\end{cases}
\end{equation}

Let $(x^a,\dot{x}^b)$ be an induced coordinate system
on the total space $T\mathcal{M}$ of the tangent bundle
$(T\mathcal{M},\Pi,\mathcal{M})$ and
in the following, where convenient, we will write $x$ instead of $x^a$ and $\dot{x}$
instead of
$\dot{x}^b$.

We are interested in the evolution of a thermal plasma over
timescales during which the motion of the ions is negligible in
comparison with the motion of the electrons. We assume that the ions are at
rest and distributed homogeneously in the laboratory frame. Their
worldlines are trajectories of the vector field $N_\text{ion} =
n_\text{ion}\partial/\partial x^0$ on $\mathcal{M}$ where
$n_\text{ion}$ is the constant ion number density measured in the laboratory
frame. The electrons are described statistically by a $1$-particle
distribution $f(x,\dot{x})$ which induces a number $4$-current vector
field $N = N^a \partial/\partial x^a$
\begin{equation}
\label{component_number_current}
N^a(x) = \int_{\mathbb{R}^3} \dot{x}^a f(x,\dot{x})\,
\frac{1}{\sqrt{1+|\dot{\bm{x}}|^2}}\, d\dot{x}^1 d\dot{x}^2 d\dot{x}^3
\end{equation}
on $\mathcal{M}$, where $|\dot{\bm{x}}|^2 = (\dot{x}^1)^2 + (\dot{x}^2)^2 +
(\dot{x}^3)^2$. One may write the Maxwell equations on $\mathcal{M}$ as
\begin{align}
\label{component_Maxwell_1}
&\frac{\partial F_{bc}}{\partial x^a} + \frac{\partial
  F_{ab}}{\partial x^c} + \frac{\partial F_{ca}}{\partial x^b} = 0,\\
\label{component_Maxwell_2}
&\frac{\partial F^{ba}}{\partial x^b} = q N^a - q N^a_\text{ion},
\end{align}
where $F_{ab}$ are the components of the electromagnetic field tensor,
$F^{ab} = \eta^{ac}\eta^{bd} F_{cd}$, $q$ is the charge
on the electron ($q<0$) and $(\eta^{ab})$ is the matrix inverse of
$(\eta_{ab})$. The scalar field $f$ satisfies the Vlasov
equation, which may be written
\begin{equation}
\label{component_vlasov}
\dot{x}^a\bigg(\frac{\partial f}{\partial x^a} - \frac{q}{m}
F^b{ }_a^{\bm{V}}\frac{\partial f}{\partial\dot{x}^b} \bigg) = 0
\end{equation}
on $T\mathcal{M}$
where $F^b{ }_a^{\bm{V}}$ is the vertical lift of $F^b{ }_a =
\eta^{bc}F_{ca}$ from $\mathcal{M}$ to $T\mathcal{M}$,
\begin{equation}
F^b{ }_a^{\bm{V}}(x,\dot{x}) = F^b{ }_a(x).
\end{equation}
\subsection*{{ Exterior formulation}}
In this section we recast the above using the tools of exterior
differential calculus as it affords a succinct and powerful language
for subsequent analysis. We make extensive use of Cartan's exterior
derivative $d$, the exterior product $\wedge$ and the Hodge map
$\star$ on differential forms (see, for example,~\cite{burton:2003,
  benn:1987}).

The spacetime volume $4$-form $\star 1$ is
\begin{equation}
\star 1 = dx^0\wedge dx^1\wedge dx^2\wedge dx^3
\end{equation}
and the Maxwell equations
(\ref{component_Maxwell_1}, \ref{component_Maxwell_2}) can be written
\begin{equation}
\label{maxwell}
dF = 0,\qquad
d\star F = - q \star \widetilde{N} + q \star \widetilde{N_{\text{ion}}}
\end{equation}
where $F =\frac{1}{2}F_{ab}\, dx^a\wedge dx^b$ is the electromagnetic
$2$-form, and the $1$-forms
$\widetilde{N}$, $\widetilde{N_{\text{ion}}}$  are the metric duals of
the vector fields $N$, $N_{\text{ion}}$ respectively. (The metric dual
$\widetilde{Y}$ of a vector field $Y$ satisfies $\widetilde{Y}(Z)=g(Y,Z)$
for all vector fields $Z$.)  

Introduce the vector fields $L$, $X$,
\begin{align}
\label{definition_L}
L &= \dot{x}^a\bigg(\frac{\partial}{\partial x^a} - \frac{q}{m}
F^b{ }_a^{\bm{V}}\frac{\partial}{\partial\dot{x}^b}\bigg),\\
\label{definition_X}
X &= \dot{x}^a\frac{\partial}{\partial\dot{x}^a},
\end{align}
on $T\mathcal{M}$ and the $6$-form $\omega$,
\begin{equation}
\label{definition_omega}
\omega = \iota_L\,\iota_X (\star 1^{\bm{V}}\wedge \# 1)
\end{equation}
on $T\mathcal{M}$ where $\iota_Y$ is the interior operator on forms
with respect to vector $Y$, the $4$-form $\star 1^{\bm{V}}$
\begin{equation}
\star 1^{\bm{V}} = dx^0 \wedge dx^1 \wedge dx^2 \wedge dx^3
\end{equation}
is the vertical lift of the spacetime volume $4$-form $\star 1$ from
$\mathcal{M}$ to $T\mathcal{M}$ and the $4$-form $\# 1$
\begin{equation}
\# 1 = d\dot{x}^0 \wedge d\dot{x}^1 \wedge d\dot{x}^2 \wedge d\dot{x}^3
\end{equation}
on $T\mathcal{M}$.

The total space $\mathcal{E}$ of the sub-bundle
$(\mathcal{E},\Pi,\mathcal{M})$ of $(T\mathcal{M},\Pi,\mathcal{M})$ is the
set of timelike, future-directed, unit normalized tangent vectors on $\mathcal{M}$,
\begin{equation}
\mathcal{E}=\{(x,\dot{x})\in
T\mathcal{M}\, \big|\, \varphi = 0\,\,\text{and}\,\, \dot{x}^0>0\}
\end{equation}
where
\begin{equation}
\varphi = \eta_{ab}\,\dot{x}^a\dot{x}^b + 1.
\end{equation}
The integral (\ref{component_number_current}) can be written 
\begin{equation}
\label{component_number_current_forms}
N^a(x) = \int_{\mathcal{E}_x} \dot{x}^a f\,\iota_X\# 1
\end{equation}
where $\mathcal{E}_x = \Pi^{-1}(x)$ is the fibre of
$(\mathcal{E},\Pi,\mathcal{M})$ over $x\in\mathcal{M}$,
and it can be shown that the Vlasov equation (\ref{component_vlasov}) can be written
\begin{equation}
d(f\omega) \simeq 0
\end{equation}
where $\simeq$ denotes equality under restriction to $\mathcal{E}$ by pull-back.  
Thus, it follows
\begin{equation}
\int_{\mathcal{B}} d(f\omega) = 0
\end{equation}
where $\mathcal{B}$ is a $6$-dimensional region in
$\mathcal{E}$ and using the generalized Stokes theorem on forms (see,
for example,~\cite{benn:1987}) we obtain
\begin{equation}
\label{stokes_on_f_omega}
\int_{\mathcal{\partial B}} f\omega = 0
\end{equation}
where $\partial\mathcal{B}$ is the boundary of
$\mathcal{B}$.
\subsection*{{ Piecewise constant distributions}}
We consider distributions for which $f=\alpha$ is a positive constant inside a
$6$-dimensional
region $\mathcal{U}\subset\mathcal{E}$ and $f=0$ outside. In
particular, we consider $\mathcal{U}$ to be the union over each point
$x\in\mathcal{M}$ of a domain $\mathcal{W}_x$ whose boundary
$\partial\mathcal{W}_x$ in $\mathcal{E}$ is
topologically equivalent to the $2$-sphere.
Such distributions are sometimes called ``waterbags'' in the
literature.

Choosing $\mathcal{B}$ in (\ref{stokes_on_f_omega}) to be a
small $6$-dimensional 
``pill-box'' that intersects $\partial\mathcal{W}_x$ and taking
the appropriate limit as the volume of $\mathcal{B}$ tends to zero, we
recover a jump condition on $f\omega$ that leads to
\begin{equation}
\label{jump_condition}
d\lambda\wedge \omega \simeq 0\,\,\,\text{at $\lambda=0$}
\end{equation}
where $\lambda=0$ is the union over $x$ of the boundaries
$\partial\mathcal{W}_x$.

If $\lambda=0$ is the image of the embedding map $\Sigma$,
\begin{eqnarray}
\notag \Sigma : \mathcal{M} \times S^2 &\rightarrow& \mathcal{E} \\
(x, \xi) &\mapsto& (x, \dot{x} = V_\xi (x)), 
\end{eqnarray}
where $\xi= (\xi^1, \xi^2)$ is a point in $S^2$, then it follows from
(\ref{definition_L}, \ref{definition_X}, \ref{definition_omega}) that
(\ref{jump_condition}) is equivalent to
\begin{equation}
 \big( \nabla_{V_\xi} \widetilde{V_\xi} - \frac{q}{m} \iota_{V_\xi} F \big) \wedge
\Omega_\xi=0.  \label{Lorentz_jump}
\end{equation}
Here, $V_\xi$ and $\Omega_\xi$ are families of vector fields and 2-forms on
$\mathcal{M}$ respectively, defined by
\begin{align}
 &V_\xi = V^a_\xi \frac{\partial}{\partial x^a}= (\Sigma^\ast \dot{x}^a)
\frac{\partial}{\partial x^a}, \\
&\Omega_\xi= \frac{\partial \Sigma^\ast \dot{x}^a}{\partial \xi^1} dx_a \wedge
\frac{\partial \Sigma^\ast \dot{x}^b}{\partial \xi^2} dx_b.
\end{align}
where $dx_a = \eta_{ab} dx^b$. 
Note that since the image of $\Sigma$ lies in $\mathcal{E}$, it
follows that, for each $\xi \in S^2$, $V_\xi$ is timelike, unit normalized and
future-directed:
\begin{align}
 \label{norm}
g(V_\xi,V_\xi) = -1,\qquad g(V_\xi, \frac{\partial}{\partial x^0}) <0.
\end{align}
We adopt (\ref{Lorentz_jump}) as the equation of motion for $\partial\mathcal{W}_x$.
 
It may be shown that a
particular class of solutions to (\ref{Lorentz_jump}) satisfies
\begin{equation}
\label{solved_Lorentz}
F = \frac{m}{q} d \widetilde{V}_\xi
\end{equation}
and using (\ref{maxwell}) we obtain the field equation
\begin{equation}
\label{maxwell_2form_eliminated}
d\star d \widetilde{V}_\xi = - \frac{q^2}{m}(\star\widetilde{N} -
\star\widetilde{N_\text{ion}})
\end{equation}
on $\mathcal{M}$ with the condition that $d\widetilde{V}_\xi$ is independent of
$\xi$. For simplicity, we have neglected the direct contribution of the laser pulse
to the total
electromagnetic field in (\ref{solved_Lorentz}).   
%
\section{Electrostatic oscillations}
Before analysing (\ref{maxwell_2form_eliminated}, \ref{norm}) further
it is useful to briefly discuss their analogue on 2-dimensional
spacetime for facilitating comparison with the approach adopted
in~\cite{katsouleas:1988}.
\subsection*{Electrostatic oscillations in 1 spatial dimension}
Although formulated on 4-dimensional spacetime, equations
(\ref{maxwell_2form_eliminated}, \ref{norm}) have a similar structure
for any number of dimensions. In particular, we now consider
2-dimensional Minkowski spacetime $(\mathcal{M},g)$
\begin{eqnarray}
g &=& - dt \otimes dt + dz \otimes dz,\\
\star 1 &=& dt \wedge dz
\end{eqnarray}
where $(t,z)$\footnote{We use $(t,z)$ rather than $(x^a)$ to
distinguish coordinates on 2- and 4-dimensional spacetimes.} is a
Cartesian coordinate system in the laboratory inertial frame. An
induced coordinate system on $T\mathcal{M}$ is
$(t,z,\dot{t},\dot{z})$ and note that in this sub-section of the
article the fibre space of
$(\mathcal{E},\Pi,\mathcal{M})$
is $1$-dimensional, whereas in the rest of the article it is
$3$-dimensional. Furthermore, $\xi$ is now an element of the
$0$-sphere $\{+,-\}$ and $\Omega_\xi = 1$ is a constant
$0$-form. Thus, the analogue to (\ref{Lorentz_jump}) is
\begin{align}
\notag
&\nabla_{V_+}
  \widetilde{V_+}
- \frac{q}{m} \iota_{V_+} F = 0,\\
\label{Lorentz_jump2d}
&\nabla_{V_-}
  \widetilde{V_-}
- \frac{q}{m} \iota_{V_-} F = 0,
\end{align}
where $V_\pm$ satisfy the conditions
\begin{align}
\notag
&g(V_+,V_+) = -1,\qquad
g(V_+, \frac{\partial}{\partial t}) <0,\\
\label{norm2d}
&g(V_-,V_-) = -1,\qquad
g(V_-, \frac{\partial}{\partial t}) <0,
\end{align}
and the only non-trivial Maxwell equation for the $2$-form $F$ is 
\begin{equation}
\label{max2d}
d \star F = - q \star \widetilde{N} + q \star \widetilde{N_{\text{ion}}}
\end{equation}
where $N_{\text{ion}} = n_{\text{ion}}\partial/\partial t$ is the ion number
$2$-current and $F = E dt\wedge dz$ where $E$ is the electric field along
the $z$-axis.

On $\mathcal{E}$, $\dot{t}= \sqrt{1+\dot{z}^2}$ and the components of
the electron number $2$-current $N=N^t \partial/\partial t + N^z \partial/\partial
z$ corresponding to
(\ref{component_number_current_forms}) are
\begin{eqnarray}
\nonumber N^t &=& \int_\mathbb{R} f(t,z,\dot{t},\dot{z})
\,d\dot{z} = \alpha \Big( X_+- X_- \Big), \\
N^z &=& \int_\mathbb{R} \frac{\dot{z}}{\sqrt{1+\dot{z}^2}} f(t,z,\dot{t},\dot{z})
\,d\dot{z}= \alpha \Big( \sqrt{1+X^2_+}- \sqrt{1+X^2_-} \Big),
\end{eqnarray}
where
\begin{equation}
f =
\begin{cases}
\alpha,\qquad X_-\le \dot{z} \le X_+,\\
0, \qquad \dot{z} < X_-\text{ or } \dot{z} > X_+
\end{cases}
\end{equation}
with $\alpha$ a positive constant and $\{X_+, X_-\}$ scalar fields
over spacetime.  The $2$-velocity fields
$\{V_+,V_-\}$ satisfy
\begin{equation}
V_\pm = \sqrt{1+X^2_{\pm}}\, \frac{\partial}{\partial t} + X_{\pm} \,
\frac{\partial}{\partial z}
\end{equation}
and it follows
\begin{equation}
 \widetilde{N}= \alpha {\star \big( \widetilde{V_+}- \widetilde{V_-} \big)}.
\end{equation}

Unlike their $4$-dimensional analogue, which may include transverse
electromagnetic fields, (\ref{Lorentz_jump2d}) are \emph{uniquely}\footnote{Proper
incorporation of transverse fields requires at
least $2$ spatial dimensions.} solved by  
\begin{equation}
d\widetilde{V_\pm} = \frac{q}{m} F \label{Lorentz2d}
\end{equation}
and using (\ref{max2d})
\begin{equation}
\label{maxwell_2form_eliminated2d}
d \star d \widetilde{V_\pm} = - \frac{q^2}{m} (\star \widetilde{N} - \star
\widetilde{N_{\text{ion}}})
\end{equation}
subject to the condition $d\widetilde{V_+} =
d\widetilde{V_-}$.

Alternatively, one may follow the approach adopted in
\cite{katsouleas:1988} employing a warm fluid model:
\begin{align}
&(\rho + p) \nabla_U \widetilde{U} = qn \iota_U F - \iota_U
  ( dp \wedge \widetilde{U} ),\\
&g(U,U) = -1,\\
\label{warm_fluid2d}
&g(U,\frac{\partial}{\partial t}) < 0.
\end{align}
Here,
\begin{equation}
N = n U
\end{equation}
where $U$ is the bulk $2$-velocity of the electron fluid,
\begin{equation}
U = \frac{1}{\sqrt{-g(Z,Z)}} Z,\qquad Z = \frac{1}{2}(V_+ + V_-),
\end{equation}
and, in the electron fluid's rest frame, $\rho$ is the fluid's energy
density and $p$ is the fluid's pressure defined as
\begin{align}
\rho = \int_\mathbb{R} \sqrt{1+\dot{z}^2}f(t,z,\dot{t},\dot{z})\,
d\dot{z}\\
p = \int_\mathbb{R} \frac{\dot{z}^2}{\sqrt{1+\dot{z}^2}}f(t,z,\dot{t},\dot{z})\,
d\dot{z}.
\end{align}
It may be shown
\begin{align}
\label{eqns_state2da}
&\rho =
  m\alpha\bigg[\frac{n}{2\alpha}\sqrt{1+\bigg(\frac{n}{2\alpha}\bigg)^2}
  + \text{sinh}^{-1}\bigg(\frac{n}{2\alpha}\bigg)\bigg],\\
\label{eqns_state2db}
&p = m\alpha\bigg[\frac{n}{2\alpha}\sqrt{1+\bigg(\frac{n}{2\alpha}\bigg)^2}
  - \text{sinh}^{-1}\bigg(\frac{n}{2\alpha}\bigg)\bigg].
\end{align}
Thus, (\ref{Lorentz_jump2d}, \ref{norm2d}) may be replaced by an equivalent field
theory expressed in terms of a finite set of moments of $f$ on
$2$-dimensional spacetime. However, the situation is more complicated
for waterbags over $4$-dimensional spacetime where the moment hierarchy is not
automatically closed. 

We will now use (\ref{maxwell_2form_eliminated2d}) to obtain a
non-linear oscillator describing 1-dimensional electrostatic oscillations.
Let all field components with respect to the laboratory frame $(dt,dz)$ be functions
of $\zeta = z - vt$ only
(the ``quasi-static assumption''), where $0<v<1$, and let $(e^1,e^2)$ be the basis
\begin{equation}
e^1= vdz-dt, \qquad e^2=dz-vdt.
\end{equation}
The coframe $(\gamma e^1,\gamma e^2)$ is an orthonormal
basis adapted to observers moving at velocity $v$ along $z$ (i.e observers in the
``wave frame'') where
$\gamma= (1-v^2)^{-1/2}$ is the Lorentz factor of such observers
relative to the laboratory. For example, $\gamma e^2(N_\text{ion}) = -
\gamma n_\text{ion} v$ is the ion $1$-current in the wave frame.

In the basis $(e^1,e^2)$, $\widetilde{V_\pm}$ can be decomposed as
\begin{equation}
\widetilde{V_\pm}= \big( \mu(\zeta) + A_\pm \big) e^1 + \psi_\pm
(\zeta) e^2.   \label{decomp2d}
\end{equation}
Note that this is the most general decomposition compatible with
equation (\ref{Lorentz2d}) and the quasi-static assumption. 

Solving (\ref{norm2d}) for $\psi^2_\pm$ gives
\begin{equation}
\psi^2_\pm = (\mu + A_\pm)^2- \gamma^2
\end{equation}
and additional physical information is needed to fix
the sign of $\psi_\pm$. Here, we demand that all electrons
described by the waterbag are
travelling slower than the wave so $\psi_\pm = -\sqrt{(\mu + A_\pm)^2- \gamma^2}$
and (\ref{decomp2d}) is 
\begin{align}
\widetilde{V_\pm} = &\big( \mu + A_\pm \big) e^1
- \Big( (\mu +A_\pm)^2- \gamma^2 \Big)^{1/2}  e^2.  
\end{align}
Substituting (\ref{decomp2d}) into equation (\ref{Lorentz2d}) yields
\begin{equation}
E= \frac{1}{\gamma^2} \frac{m}{q} \frac{d\mu}{d\zeta},
\end{equation}
and equation (\ref{maxwell_2form_eliminated2d}) yields the nonlinear oscillator
equation
\begin{equation}
\label{osc2d}
\frac{1}{\gamma^2} \frac{d^2 \mu}{d\zeta^2} = -\frac{q^2}{m} \gamma^2 n_\text{ion} -
\frac{q^2}{m} \alpha 
\bigg[ \sqrt{(\mu + A_+)^2 -\gamma^2} -\sqrt{(\mu + A_-)^2 -\gamma^2} \bigg]
\end{equation}
with the algebraic constraint
\begin{equation}
\label{alg_const2d}
A_+ -A_- = -\frac{n_\text{ion} \gamma^2 v}{\alpha}.
\end{equation}

\subsection*{Longitudinal electrostatic oscillations in 3 spatial dimensions}
We now consider electrostatic waves in 3 spatial dimensions by closely
following the above description of 1 dimensional electric waves.  

To proceed further we seek a form for $\mathcal{W}_x$ axisymmetric
about $\dot{x}^3$ whose pointwise dependence in 
$\mathcal{M}$ is on the wave's phase $\zeta = x^3 - v x^0$ only, where
$0<v<1$. As before, the following results are applicable only if the
longitudinal component of $V_\xi$ in the wave frame is negative
(no electron described by $\mathcal{W}_x$ is moving faster along $x^3$
than the wave).

Decompose $\widetilde{V}_\xi$ in the wave frame as
\begin{equation}
\widetilde{V}_\xi = [\mu(\zeta) + A(\xi^1)]\, e^1 + \psi(\xi^1,\zeta)\, e^2
\,\,+ R\sin(\xi^1)\cos(\xi^2)dx^1 + R\sin(\xi^1)\sin(\xi^2)dx^2  \label{V_ansatz}
\end{equation}
for $0 < \xi^1 < \pi$, $0 \le \xi^2 < 2\pi$ 
where $R>0$ is constant
and
\begin{equation}
\label{coframe}
e^1 = v dx^3 - dx^0,\qquad e^2 = dx^3 - v dx^0.
\end{equation}
Here, $(\gamma e^1, \gamma e^2, dx^1, dx^2)$ is an orthonormal basis (the wave
frame) with $\gamma = 1/\sqrt{1-v^2}$.
In the wave frame the relativistic energy of $P_\xi = m V_\xi$ is $m(\mu +
A)/\gamma$ 
and it
follows that $\mu+A > 0$. The component $\psi$ is determined using (\ref{norm}),
\begin{equation}
\label{psi}
\psi = -\sqrt{[\mu + A]^2 - \gamma^2[1 + R^2 \sin^2(\xi^1)]},
\end{equation}
where the negative square root is chosen because no electron is moving
faster along $x^3$ than the wave. 
 
Substituting (\ref{V_ansatz}) into equation (\ref{solved_Lorentz}) leads to 
\begin{equation}
\label{F_dmu}
F= \frac{m}{q} \frac{d \mu}{d\zeta} e^2 \wedge e^1,
\end{equation}
and (\ref{maxwell_2form_eliminated},
\ref{component_number_current_forms}, \ref{V_ansatz}, \ref{psi}) yield
\begin{equation}
\frac{1}{\gamma^2}\frac{d^2\mu}{d\zeta^2} = -
  \frac{q^2}{m}n_{\text{ion}}\gamma^2
- \frac{q^2}{m}2\pi R^2 \alpha \int\limits^\pi_0 \bigg([\mu +
    A(\xi^1)]^2
\,\,- \gamma^2[1 + R^2
    \sin^2(\xi^1)]\bigg)^{1/2}\sin(\xi^1)\,\cos(\xi^1)\, d\xi^1   \label{ODE_mu}
\end{equation}
(c.f. equation (\ref{osc2d})) and
\begin{equation}
\label{norm_A}
2\pi R^2 \int\limits^\pi_0
A(\xi^1)\,\sin(\xi^1)\,\cos(\xi^1)\,d\xi^1 = - \frac{n_\text{ion}\gamma^2\,v}{\alpha}
\end{equation}
(c.f. equation (\ref{alg_const2d})) where $\alpha$ is the value of $f$ inside
$\mathcal{W}_x$.

The form of the 2nd order autonomous
non-linear ordinary differential equation (\ref{ODE_mu}) for $\mu$ is fixed by
specifying the generator $A(\xi^1)$ of $\partial\mathcal{W}_x$ subject to the
normalization condition (\ref{norm_A}).
\section{Electrostatic wave-breaking}
The form of the integrand in (\ref{ODE_mu}) ensures that the magnitude
of oscillatory solutions to (\ref{ODE_mu}) cannot be arbitrarily
large. For our model, the wave-breaking value $\mu_{\text{wb}}$
is the largest $\mu$ for which the argument of the square root in
(\ref{ODE_mu}) vanishes,
\begin{equation}
\mu_{\text{wb}} =
\text{max}\bigg\{-A(\xi^1) + \gamma\sqrt{1+R^2\sin^2(\xi^1)}
\,\bigg|\,0\le\xi^1\le\pi\bigg\},  \label{mu_wave-breaking}
\end{equation}
because $\mu<\mu_{\text{wb}}$ yields an imaginary integrand in
(\ref{ODE_mu}) for some
$\xi^1$. The positive square root in (\ref{mu_wave-breaking}) is
chosen because, as discussed above, $\mu + A(\xi^1) > 0$ and in
particular $\mu_\text{wb} + A(\xi^1)>0$.

The electric field has only one non-zero component $E$ (in the $x^3$
direction). Using $F=E\,dx^0\wedge dx^3$ and (\ref{V_ansatz},
\ref{coframe}, \ref{F_dmu}) it follows
\begin{equation}
\label{E_dmu}
E = \frac{m}{q} \frac{1}{\gamma^2} \frac{d\mu}{d\zeta}
\end{equation}
and the wave-breaking limit $E_{\text{max}}$ is obtained by evaluating
the first integral of (\ref{ODE_mu}) between $\mu_{\text{wb}}$ where
$E$ vanishes and the 
equilibrium\footnote{\label{footnote1}Note that the equilibrium of
  $\mu$ need not coincide with the plasma's thermodynamic equilibrium.} value
$\mu_{\text{eq}}$ of $\mu$ where $E$ is at a maximum. Using
(\ref{norm_A}) to eliminate $\alpha$ it follows that $\mu_{\text{eq}}$ satisfies
\begin{equation}
\frac{1}{v}\int\limits^\pi_0 A(\xi^1)\sin(\xi^1)\cos(\xi^1)\,d\xi^1
= \int\limits^\pi_0 \bigg([\mu_{\text{eq}} + A(\xi^1)]^2
- \gamma^2[1 + R^2
    \sin^2(\xi^1)]\bigg)^{1/2}
\sin(\xi^1)\cos(\xi^1)
    d\xi^1   \label{mu_equilibrium}
\end{equation}
with 
\begin{equation}
\label{A_negativity}
\int\limits^\pi_0 A(\xi^1)\sin(\xi^1)\cos(\xi^1)\,d\xi^1\, <\, 0
\end{equation}
since $\alpha, v >0$. Equation (\ref{ODE_mu}) yields the
maximum value $E_\text{max}$ of $E$,
\begin{align}
\notag E_{\text{max}}^2 = 2 m n_\text{ion}\Bigg[
-\mu_{\text{eq}} + \mu_{\text{wb}}
+ \, &\frac{v}{\int\limits^\pi_0
    A(\xi^{1\prime})\sin(\xi^{1\prime})\cos(\xi^{1\prime})d\xi^{1\prime}} \times \\
& \int\limits^{\mu_{\text{eq}}}_{\mu_{\text{wb}}}\int\limits^\pi_0
\bigg([\mu + A(\xi^1)]^2
- \gamma^2 [1 + R^2
    \sin^2(\xi^1)]\bigg)^{1/2}
\sin(\xi^1)\cos(\xi^1)
    d\xi^1\,d\mu\Bigg].   \label{E_max}
\end{align}


\subsection*{{ Example}}
The above may be used to determine a
wave-breaking limit for a nearly cold plasma whose distribution's
transverse extent is much larger than its longitudinal extent.   

Let $A(\xi^1) = -a\cos(\xi^1)$ where $a$ is a positive constant.
Using (\ref{E_max}) it follows
\hspace{-2em}
\begin{equation}
E_{\text{max}}^2 = 2 m n_\text{ion}\Bigg[
-\mu_{\text{eq}} + \mu_{\text{wb}}
+ \frac{3}{2}\frac{v}{a}
\int\limits^{\mu_{\text{eq}}}_{\mu_{\text{wb}}}\int\limits^1_{-1}
\bigg([\mu + a\chi]^2
- \gamma^2[1 + R^2 (1-\chi^2)]\bigg)^{1/2}\chi\,d\chi\,d\mu
\Bigg]   \label{E_max_example}
\end{equation}
where $\chi=-\cos(\xi^1)$ and equation (\ref{mu_equilibrium}) yields
\begin{equation}
\frac{3}{2}\frac{v}{a}\int\limits^1_{-1} \bigg([\mu_{\text{eq}} +
  a\chi]^2
- \gamma^2[1 + R^2(1-\chi^2)]\bigg)^{1/2}\chi\,d\chi = 1.  
\label{mu_equilibrium_example}
 \end{equation}
Equation (\ref{mu_wave-breaking}) may be written
\begin{equation}
\mu_{\text{wb}} =
\text{max}\bigg\{-a\chi + \gamma\sqrt{1+R^2(1-\chi^2)}
\,\bigg|\,-1\le\chi\le 1\bigg\},   \label{mu_wave-breaking_max_example}
\end{equation}
and for $a,R,\gamma$ satisfying
\begin{equation}
\frac{a}{R}\sqrt{\frac{1+R^2}{a^2+\gamma^2 R^2}} < 1,\qquad R > 0
\end{equation}
the maximum of $h(\chi)=-a\chi + \gamma\sqrt{1+R^2(1-\chi^2)}$ over 
$-1\le\chi\le 1$ coincides with a turning point $\chi =
\hat{\chi}=-\cos(\hat{\xi}^1)$ of $h$
where
\begin{equation}
\cos(\hat{\xi}^1) = \frac{a}{R}\sqrt{\frac{1+R^2}{a^2+\gamma^2 R^2}}.
\end{equation}
During the maximum amplitude oscillation the points
$\xi^1 = \hat{\xi}^1$ catch up with the wave and it follows
\begin{equation}
\label{mu_wave-breaking_example}
\mu_{\text{wb}} = \frac{1}{R}\sqrt{(1+R^2)(a^2+\gamma^2 R^2)}.
\end{equation}
For $a\ll R\ll 1$ equations (\ref{E_max_example}, \ref{mu_equilibrium_example},
\ref{mu_wave-breaking_example}) yield
\begin{equation}
E_\text{max}^2 \approx \frac{2m^2c^2\omega_p^2}{q^2} \bigg(\gamma - 1 - \frac{3}{4}
\frac{v}{c}\gamma R\bigg)
\end{equation}
where $m c \omega_p \sqrt{2(\gamma - 1)}/|q|$ is the
usual relativistic cold plasma wave-breaking
limit of $E$ (see, for example,~\cite{mori:1990}) and
$\omega_p=\sqrt{n_\text{ion}q^2/(m\varepsilon_0)}$ is the plasma
angular frequency. Note that the speed
of light $c$ and the permittivity $\varepsilon_0$ of the vacuum have
been restored. The parameter $R$ may be
eliminated in favour of an effective transverse ``temperature''
$T_{\perp\text{eq}}$ defined as 
\begin{align}
&T_{\perp\text{eq}} = \frac{1}{2k_B\,n_{\text{ion}}}(P^{11}_\text{eq}
  + P^{22}_\text{eq}),\\
&P^{ab}_\text{eq} = m\alpha \int_{\mathcal{W}_{\text{eq}}}
  \dot{x}^a\dot{x}^b
  \iota_X \# 1
\end{align}
where $\mathcal{W}_{\text{eq}}$ is the support of the distribution
with $\mu=\mu_{\text{eq}}$ (see footnote~\ref{footnote1}) and $k_B$ is
Boltzmann's constant. It follows
\begin{equation}
R \approx \sqrt{\frac{5k_B T_{\perp\text{eq}}}{m c^2}}
\end{equation}
where the speed of light $c$ has been restored.
\section*{Conclusion}
We have developed a method for investigating the relationship between the
shape of a $1$-particle distribution and electrostatic non-linear
thermal plasma waves near breaking. An approximation to the
wave-breaking limit of the electric field was obtained for a particular
axisymmetric distribution.  

Further analysis of (\ref{E_max}, \ref{mu_equilibrium},
\ref{mu_wave-breaking}) will be presented elsewhere.
\section*{Acknowledgements}
We thank RA Cairns, B Ersfeld, A Reitsma and RMGM Trines
for useful discussions. This work
is supported by EPSRC grant EP/E022995/1.  

\end{document}